\documentstyle[12pt]{article}
\begin{document}
\hsize 6.5 in
\oddsidemargin 0in
\baselineskip 14pt
\vspace* {.1in}
{\large \bf \centerline {SEEING PLANCK SCALE PHYSICS AT ACCELERATORS}}

\vspace{0.3in}

{\large 
\centerline {\it R. Arnowitt $^{1}$ and Pran Nath $^2$}
\vspace{0.3in}
\centerline{$^1$Center for Theoretical  Physics, Department of Physics }
\centerline{Texas A\&M University, College Station, TX 77843-4242}
\centerline {$^2$Department of Physics, Northeastern University,}
\centerline {Boston, MA 02115}}

\vspace{.2in}

{\small

\centerline  {\bf Abstract}

Much current theoretical analysis is based on the hypothesis that the physics
beyond the Standard Model is a consequence of new principles that occur at the 
Planck scale. The question arises whether such principles can ever be directly
tested. We show here that for a significant class of models, hypotheses at the
string or Planck scale can indeed  be directly tested to relatively high precision
by linear colliders. Three classes of models are examined: those with universal
SUSY soft breaking at the string scale; those with a horizontal symmetry at the
string or Planck scale, and simple Calabi-Yau superstring models with dilaton
and moduli SUSY breaking.}

\vspace{.2in}

\noindent
{\bf INTRODUCTION}

\vspace{.2in}

While the Standard Model is a remarkably succesful theory, heaving been subjected
to numerous high precision experimental tests, there are many aspects about it 
that are not understood, e.g. Yukawa couplings, CKM parameters, etc.  
There has been much theoretical analysis based on the assumption that these 
items are a consequence of new physical principles arising at or near the Planck 
scale,  $M_{Pl}\cong 2.4 \times 10^{18}\ GeV$. 
The strongest direct evidence that the ultra high energy domain may
influence low energy phenomena resides in the success of supersymmetric (SUSY) grand
unification, i.e. that the three coupling constants $\alpha _1$, $\alpha_2$, $\alpha_3$
unify at a GUT scale $M_{G}\cong 2 \times 10^{16}\ GeV$ to a value $\alpha_{G}\cong 
1/24$ for a minimal SUSY particle spectrum with one pair of Higgs doublets. 
First observed in the 1990 LEP data \cite{lang}, this result has stood the test of time, 
both with refinements in the data and refinements in the theoretical treatment [including
for the latter SUSY threshold effects at $M_S \cong 100\ GeV-1\ TeV$, GUT scale threshold 
effects near $M_G$, and possible small Planck physics effects 
from non-renormalizable operators (NRO)].

We will use supergravity (SUGRA) grand unification \cite{cham} to analyse these questions here. 
There has been much discussion over the past year as to at what
scale SUSY breaks in such models. 
We will assume here that supersymmetry breaks 
in a hidden sector at a scale above $M_G$, e.g. at O($M_{Pl}$). 
These models have a number of undetermined aspects presumably representing 
Planck scale physics, and 
we will examine three types of post-GUT assumptions that might be made:

1. Models with universal SUSY soft breaking masses at the string scale  
$M_{str}\cong  5\times 10^{17}\ GeV$.

2. Models with SU(2)$_{H}$ horizontal symmetry.

3. String models with Calabi-Yau compactification.

We will see in the above examples that linear colliders have in fact the ability to probe  physics in the post-GUT region  up to the string or Planck scales to a 
remarkable degree of precision, distinguish between different models and 
actually measure parameters that are expected to be predictions of string models.

\vspace{.2in}
\noindent
{\bf SUPERGRAVITY MODELS}

\vspace{.2in}

We review briefly the basic elements of supergravity models. 
These models depend in general on three functions of the scalar fields 
\{$\phi_i$\} (squarks, sleptons, etc.): $f_{\alpha \beta}(\phi _i)$, 
the gauge kinetic function, $K(\phi _i, \phi _i^{\dag})$,
the Kahler potential and $W(\phi _i)$, the superpotential. 
$f_{\alpha \beta}$ modifies the gauge and gaugino kinetic energies 
(e.g. $f_{\alpha \beta} F_{\mu \nu}^{\alpha}F^{\mu \nu \beta}$, 
$\alpha , \beta =$ gauge indices).  
$K$ enters into the scalar and chiral partner kinetic energies 
(e.g. $K_j^i \partial_{\mu}\phi _i \partial^{\mu} \phi _j^{\dag}$ 
where $K_j^i\equiv \partial^2 K/(\partial \phi _i \partial \phi _j^{\dag})$ and elsewhere. 
$W$ and $K$ enter in the Lagrangian only in the combination
\begin{equation}
G(\phi_i, \phi_i^{\dag})\ =\  \kappa^2K(\phi_i, \phi_i^{\dag})+ln[\kappa^6\ |W(\phi _i)|^2]
\end{equation}
where $\kappa \equiv 1/M_{Pl}$. 
To maintain the gauge hierarchy we assume the superpotential decomposes into 
a "physical" and a "hidden" sector,
\begin{equation}
W(\phi _i)\ =\ W_{phys}(\phi _a)\ +\ W_{hid}(z)
\end{equation}
Here $\{\phi_i\}=\{\phi_a,z\}$ where $\phi_a$ are physical fields and $z$ are 
fields whose VEVs spontaneously break supersymmetry and obey 
$\langle z\rangle  = O(M_{Pl})$ and $\kappa^2\langle W_{hid}\rangle = O(M_S)$. 

The mass dimensions of the basic functions are $[f_{\alpha \beta}] = $(mass)$^0$,
$[K] = $(mass)$^2$, and $[W] = $(mass)$^3$. It is convenient to introduce the
dimensionless variables $x\equiv \kappa z$ with $\langle x\rangle \ =\ O(1)$ and 
expand $f_{\alpha \beta}$, $K$ and $W$ in powers of $\phi_a$ with higher 
terms scaled by $\kappa$. 
Thus one can write

\begin{equation}
f_{\alpha \beta}(\phi_i)=c_{\alpha \beta}(x)
+\kappa c_{\alpha \beta}^a(x)\phi _a+...
\end{equation}

\begin{equation}
K(\phi _i, \phi _i^{\dag})
=\kappa ^{-2}c(x, x^{\dag})+c_b^a\phi _a\phi _b^{\dag}
+(c^{ab}\phi _a\phi _b+c_{ab}\phi _a^{\dag}\phi _b^{\dag})
+\kappa c_{bc}^a \phi_a \phi_b^{\dag} \phi_c^{\dag} + ...
\end{equation}

\begin{equation}
W_{phys}(\phi _i)={1\over6}\lambda^{abc}\phi_a\phi_b\phi_c
+{1\over24}\kappa \lambda^{abcd}\phi_a\phi_b\phi_c\phi_d+...
\end{equation}

The assumption that after SUSY breaking, the VEVs of $c_{\alpha \beta}$, $c$, 
$c_b^a$ etc. are all of $O(1)$, implies that the higher order terms scaled by 
$\kappa$ are of $O(1/M_{Pl})$ and presumably represent Planck scale physics corrections (e.g. arising in string theory from integrating out 
the towers of Planck mass states). 
The terms  with dimensionless coupling constants are accessible to low 
energy physics  (e.g. $\lambda^{abc}$ are Yukawa couplings). 
The holomorphic terms in $K$ ($c^{ab}\phi_a\phi_b$ etc.) can be transferred to 
$W_{phys}$ by a Kahler transformation,
and then give rise naturally to a $\mu$ term in $W_{phys}$ after 
SUSY breaking of size $\mu=O(M_S)$ \cite{soni}.

The spontaneous breaking of supersymmetry gives rise to the SUSY soft 
breaking masses  \cite{cham,barbi}. 
For SUSY breaking above $M_G$, the pattern of soft breaking
masses must obey the symmetries of the GUT group ${\cal G}$. 
We consider here the cases where ${\cal G}$ contains an $SU(5)$ subgroup 
(e.g. $SU(N)$, $N>5$; $SO(N)$, $N>10$; $E_6$)
and label the light matter at $M_G$ by their $SU(5)$ quantum numbers. 
Thus for three
generations of $10$ and $\bar{5}$ representations labeled by $a=1,2,3$ and one 
pair of Higgs (${\cal H}_1=\bar{5}, {\cal H}_2=5$) one has

$$
10_a=\{q_a=(\tilde{u}_{La}, \tilde{d}_{La}); \quad 
u_a\equiv \tilde{u}_{Ra}; \quad e_a\equiv \tilde{e}_{Ra}\}
$$
\begin{equation}
\bar{5}_a=\{l_a\equiv(\tilde{\nu}_{La}, \tilde{e}_{La}); \quad  
d_a\equiv \tilde{d}_{Ra}\}
\end{equation}
(For the flipped $SU(5)$ model \cite{ant}, one interchanges $\tilde{u}$ and 
$\tilde{d}$, $\tilde{\nu}$ and $\tilde{e}$ with $\tilde{e}_R$ appearing in an 
extra $SU(5)$ singlet.)
Each representation can have an independent soft breaking mass which we 
parametrize as

$$
m_{10_a}^2=m_0^2(1+\delta_a^{10});\ \ \ m_{\bar{5}_a}^2=m_0^2(1+\delta_a^{\bar{5}})
$$
\begin{equation}
m_{H_1}^2=m_0^2(1+\delta_1);\ \ \ m_{H_2}^2=m_0^2(1+\delta_2)
\end{equation}

\noindent
where $m_0=O(M_S)$. Thus a general model of this type can have eight soft 
breaking scalar masses, though specific models may have fewer parameters. 
(In addition there are, of course, gaugino masses and cubic and quadratic soft 
breaking parameters.) 
The $\delta_i$ measure the amount non-universality in the scalar soft 
breaking masses. 
One has at the grand unification scale $M_G$ the 15 relations:

\begin{equation}
\delta_{qa}=\delta_{ua}=\delta_{ea}=\delta_a^{10}; \quad 
\delta_{la}=\delta_{da}=\delta_a^{\bar{5}}
\end{equation}

Below $M_G$, the standard model gauge group holds for many models. 
One may make
contact with low energy physics by running the renormalization group equation 
(RGE) from $M_G$ down to the electroweak scale $M_Z$. 
Remarkably, the spontaneous breaking of supersymmetry at $M_G$ triggers 
the breaking of $SU(2)\times U(1)$ at
low energy \cite{kin,iban}, the scale at which electroweak breaking occurs 
being determined in large part by the top quark mass, which we take here to be $m_t=175\ GeV$.

Similarly, one may try to make contact with Planck scale physics by running 
the RGE upwards to scales above $M_G$. Here, however, results depend upon 
the particle spectrum and gauge group ${\cal G}$ above $M_G$. 
We will see that linear colliders are sensitive to both these types of
model dependences and hence will be able to distinguish between 
different possibilities.

\vspace{.2in}

\noindent
{\bf STRING SCALE UNIVERSALITY}

\vspace{.2in}

Non-universal soft breaking masses at $M_G$  can arise from running the RGE 
down from a higher scale, the non-universal effects being due to different 
Yukawa couplings etc. 
We consider in this section the case where all soft breaking masses are 
universal at the string scale $M_{str}$ \cite{polo}. 
Above $M_G$, the gauge group ${\cal G}$ is unbroken and different gauge groups 
will give different results.

\vspace{.1in}

(i) ${\cal G}=SU(5)$

\vspace{.1in}

We restrict the discussion to third generation effects, 
which have the largest Yukawa couplings, and assume that above $M_G$ 
there is $10+\bar{5}$ of the matter 
and a $5+\bar{5}+24$ of Higgs representations present.  
The superpotential has the form
\begin{equation}
W_{phys}=[h_t(10)(10)H_5+h_b(10)(\bar{5})H_{\bar{5}}]
+[Mtr(24)^2+\lambda_1tr(24)^3+\lambda_2H_5(24)H_{\bar{5}}
+\mu_0H_5H_{\bar{5}}]
\end{equation}
where $M=O(M_G)$. 
One may chose the reference mass $m_0$ to be $m_{10}$, and then
one has the following for the remaining scalar soft breaking masses:
\begin{equation}
m_{\bar{5}}^2=m_0^2(1+\delta_5); 
\quad  m_{H_{1,2}}^2=m_0^2(1+\delta_{1,2})
\end{equation}
The values of $\delta_5$, $\delta_{1,2}$ at $M_G$ are model dependent, 
depending on the coupling constants and interactions chosen in Eq. (9). 
Characteristically one finds from running the RGE from $M_{str}$ to 
$M_G$ that $\mid \delta_i \mid 
{\tiny \begin{array}{l}  < \\ \sim \end{array}} 1/2$. 
We will in the following assume
\begin{equation}
-1\leq \delta_i \leq  1
\end{equation}
The values of $m_0$, $\delta_5$, $\delta_{1,2}$ are not determined by 
supergravity theory, but presumably will be set by future Planck scale physics. 
However, they can be experimentally measured by using the RGE and 
relating them to electroweak scale quantities. 
Thus one has \cite{iban,arn}
\begin{equation}
m_0^2=m_{\tilde{e}_R}^2 - 0.151m_{1/2}^2-\sin^2\theta_W M_Z^2 \cos 2\beta
\end{equation}
\begin{equation}
m_0^2\delta_5=m_{\tilde{e}_L}^2-m_{\tilde{e}_R}^2
-0.377m_{1/2}^2+({1\over 2}-\sin^2\theta_{W})M_Z^2
\cos 2\beta
\end{equation}
where $\tilde{e}_{R,L}$ are the $R,L$ selectrons, 
$m_{1/2}=(\alpha_G/\alpha_2)\tilde{m}_2$ 
where $\tilde{m}_2$ is the $SU(2)$ gaugino mass, 
and $\tan \beta=\langle H_2\rangle /\langle H_1 \rangle$. 
The numerics in the above formulae
arise from running the RGE from $M_G$ to the electroweak scale. 
Studies have been made as to how accurately these parametars can be 
measured at the proposed Next Linear Collider (NLC) [10-13]. 
Thus it is expected that $m_{\tilde{e}_{R,L}}$ could be measured to 
1\% accuracy, $\tilde{m}_2$ to 3\%, $\tan \beta$ to 10\% and $\alpha_G$ 
perhaps to 3\%. 
Suppose, for example, actual measurements at the NLC found 
$m_{\bar{e}_L}=240 GeV$,  
$\tilde{m}_2=120\ GeV$ and $\tan \beta=5$ with the above accuracies.  
Then Eqs. (12, 13) imply \cite{arn}
\begin{equation}
m_0=(187\pm 3)GeV;\ \ \ \delta_5=0.206\pm 0.031
\end{equation}
The above discussion shows that the value of $m_0$ and the existance of 
non-universal soft breaking (i.e. $\delta_5\neq 0$) can be established at the 
NLC to remarkable accuracy. 
Further, there are many other ways of measuring $m_0$ and $\delta_5$, 
e.g. via squark masses which would act as a check on the validity of the 
model, and allow one to reduce the experimental errors.

The parameters $\delta_1$ and $\delta_2$ could be determined from 
$\mu$ and $m_A$,  where $A$ is the CP odd neutral Higgs boson. 
One has \cite{iban,arn}
$$
\mu^2(t^2-1)=[\delta_1- {1\over2}t^2(1+D_0)\delta_2]m_0^2+[1-{1\over2}t^2(3D_0-1)]m_0^2
$$
\begin{equation}
+[0.528+t^2(3.22-3.80D_0+0.060D_0^2)]m_{1/2}^2
+{1\over2}t^2(1-D_0){A_{R}^2\over D_0}
-{1\over2}M_{Z}^2(t^2-1)
\end{equation}
$$
m_A^2{t^2-1\over t^2+1} = \bigl[ \delta_1-{1\over2}(1+D_0)\delta_2
+{3\over2}(1-D_0)\bigr] m_0^2
$$
\begin{equation}
+[3.22-3.80D_0
+0.060D_0^2]m_{1/2}^{2}+{1\over2}(1-D_0){A_R^2\over D_0}
-{t^2-1\over t^2+1}M_Z^2
\end{equation}
where $D_0\cong 1-(m_t/200 \sin \beta)^2$, 
$A_R\cong A_t-0.613 m_{\tilde{g}}$, $\tilde{g}$ is the gluino and $A_t$ is the 
t-quark cubic soft breaking parameter at the electroweak scale. 
($D_0=0$ is the t-quark Landau pole and $A_R$ is the residue at the pole). 
For a linear collider (LC) of sufficient energy that heavy neutralinos 
($\tilde{\chi}_{3,4}$) can be produced \cite{fen} and A pair produced \cite{hab}, 
both $\mu$ and $m_A$ can be determined to about 2\%. 
One expects $A_R$ would 
have an error of about 5\%. 
For example, if measurements were to yield $m_0=200\ GeV$, $\mu=325\ GeV$, 
$m_A=400\ GeV$, one finds from Eqs. (15,16) that the GUT scale parameters are
\begin{equation}
m_{H_1}=(256\pm 15)\ GeV;\ \ \ m_{H_2}=(144\pm 35)\ GeV
\end{equation}
and hence $\delta_1=0.634\pm 0.220$ and $\delta_2=0.485\pm 0.178$, 
allowing a test of whether the Higgs masses are universal at $M_G$.

Further, from Eq. (8) there are three differences where this model predicts 
that the non-universal soft breaking effects should cancel out:
\begin{equation}
m_{\tilde{u}_L}^2-m_{\tilde{u}_R}^2; \quad 
m_{\tilde{u}_L}^2-m_{\tilde{e}_R}^2; \quad 
 m_{\tilde{e}_L}^2-m_{\tilde{d}_R}^2
\end{equation}
Finally we note that if the model is correct, one can use the RGE and run 
all the scalar soft breaking masses up to $M_{str}$ where they all should 
approach a common value $(m_0)_{str}$. 
Thus one can even determine in this way the value of $M_{str}$ 
experimentally, i.e. it would be the scale at which the masses unify.

\vspace{.1in}

(ii) ${\cal G}=SO(10)$

\vspace{.1in}

One can carry out a similar analysis for the $SO(10)$ group. 
We consider here the case where $SO(10)$ breaks directly to the 
Standard Model at $M_G$ and the $5+\bar{5}$
of $SU(5)$ Higgs both reside in the same 10 of $SO(10)$. 
Under these circumstances, all the $SU(5)$ relations 
considered above still hold and in addition there is the constraint
\cite{kaw} $\delta_5=\delta_2-\delta_1$. 
For the parameters discussed above one finds
\begin{equation}
\delta_5/(\delta_2-\delta_1)=0.160\pm 0.040
\end{equation}
which would imply for this case that the $SO(10)$ relation is strongly violated. 
Thus one could experimentally distinguish between different gauge groups.

\vspace{.1in}

(iii) Distinguishing Between Different Post-GUT Groups

\vspace{.1in}

Much of the physics below the GUT scale is insensitive to the nature of the 
GUT group that holds above $M_G$. 
However, the above analysis shows that linear colliders should
be able to give information about physics beyond $M_G$, and distinguish 
between different GUT groups. 
Thus if, experimentally one finds that $\delta_{u_a}\neq \delta_{e_a}$ and $\delta_{u_a}$, $\delta_{e_a}\neq 0$, and if $\delta_{d_a}\neq \delta_{l_a}$, 
then the SM gauge group would be valid, and $SU(5)$ or $SO(10)$ 
would be eliminated. 
If, however, one finds $\delta_{u_a}=\delta_{e_a}=0$ and 
$\delta_{d_a}=\delta_{l_a}$ then either $SU(5)$ or $SO(10)$ could be valid 
above $M_G$, and one could distinguish between them by checking whether 
the relation $\delta_5=\delta_2-\delta_1$ holds. 
The size of the soft breaking parameters, and generation dependence could 
give additional information about the post-GUT physics.

\vspace{.2in}

\noindent
{\bf HORIZONTAL SYMMETRIES}

\vspace{.2in}

One of the important and unresolved problems in the SM is the hierarchy of 
quark and lepton masses. 
This is related to the problem of suppressing flavor changing neutral currents 
(FCNC) arising at the loop level from both quarks and squark interactions.
An interesting approach to these questions involves imposing an $SU(2)_H$ 
horizontal symmetry in generation space \cite{dine}. 
Here one puts the first two generations into an $SU(2)_H$ dublet, 
and the third generation into an $SU(2)_H$ singlet. 
We consider here the case where the GUT group is chosen to be \cite{arn} 
${\cal G}=SU(5)\times SU(2)_H$. 

In such models, one assumes that $SU(2)_H$ is broken by $SU(2)_H$ doublet 
Higgs fields $\phi_i$ whose VEV splits the quark and lepton masses in the 
first two generations. 
In order to get the experimental pattern of quark and lepton masses, 
one requires \cite{dine} $\epsilon = \langle \phi_i \rangle/M_{Pl} \approx 1/10$. 
The picture one has, then, is that supersymmetry breaks at Planck scale, 
$SU(2)_H$ at the string scale ($M_{str}\cong M_{Pl}/10$) 
and finally $SU(5)$ at the GUT scale ($M_G\cong M_{str}/10$), 
and we will assume this in the following.

The breaking of $SU(2)_H$ produces an $O(\epsilon^2)$ splitting in 
the first two squark and slepton generations. 
This is small and thus helps to suppress FCNC in the $K^0\rightarrow \bar{K}^0$ 
and $K_L\rightarrow \mu^+ \mu^-$ SUSY box diagrams.  
Thus  neglecting this $O$(1\%) effect, one has at $M_G$ the following pattern 
of scalar soft breaking masses: $(m_{i\bar{5}})^2=m_0^2(1+\delta_5^d)$; 
$M_{10}^2=m_0^2(1+\delta_{10}^s)$; $m_{\bar{5}}^2=m_0^2(1+\delta_5^d)$; 
$m_{H_{1,2}}^2=m_0^2(1+\delta_{1,2})$ where $i=1, 2$ is a generation index, 
and masses without this label are singlet third generation masss. 
In the above, we have taken the first two generation masses of the 
10 representation as the reference mass i.e. $m_0=m_{i10}$. 
The above model thus depends on six mass parameters.

While the small splitting of the doublet degeneracy will be difficult to measure 
directly, much of the other structure will be accessible to a LC. 
There are eight independent sfermion mass measurements which can be used to 
determine the singlet parameter $\delta_{10}^s$ and four that can be used to 
determine the doublet singlet splitting $\delta_5^d-\delta_5^s$. 
Thus, if each measurement is accurate to 15\%, and the model were valid, one 
could determine each parameter to (5-10)\% 
accuracy, giving a reasonable test 
of the model. 
Further, unlike models which assume universality at $M_{str}$, using the RGE to 
proceed to scales above $M_G$ would not be expected to lead to the doublet and 
singlet soft breaking masses unifying at the higher string scale, since the third generation is in a different $SU(2)_H$ representation from the first two. 
Thus this model is distinguishable from those of the previous section.

\vspace{.2in}

\noindent
{\bf SUPERSTRING MODELS}

\vspace{.2in}

The mechanism for supersymmetry breaking in superstring theory is not 
understood at present, and as a consequence one cannot make phenomenological predictions in string theory from first principles. 
However, it has been suggested that SUSY breaking in string theory may arise 
from VEV formation of the dilaton field and T and U moduli fields \cite{font}. 
We consider here some simple Calabi-Yau models where ${\cal G}=E_6\times E_8$ 
with only a single T modulus \cite{can}. 
For these cases, the soft breaking masses arising from dilaton and T moduli VEVs 
are actually universal at $M_{str}$. 
However, the string theory imposes additional constraints which can be 
experimentally tested at linear colliders. 
Thus one has the following relations at $M_{str}$ \cite{kim}.
\begin{equation}
m_{1/2}=\sqrt{3}\sin \theta\ e^{-\imath \gamma_S}m_{3/2}
\end{equation}
\begin{equation}
m_0^2=[\sin^2\theta +\cos^2 \theta\ \triangle(T,T^*)]m_{3/2}
\end{equation}
\begin{equation}
A_0=-\sqrt{3}[\sin \theta\ e^{-\imath \gamma_S}
+\cos \theta\ e^{-\imath \gamma_T}\omega (T,T^*)]m_{3/2}
\end{equation}
where $m_{3/2}$ is the gravitino mass, $m_{1/2}$ is the universal gaugino mass, 
$A_0$ is the universal cubic soft breaking parameter, 
$\gamma_{S,T}$ are possible PC violating phases, 
and $\theta$ is the dilaton-Goldstino angle. 
$\triangle$ and $\omega$ include $\sigma$ model corrections and instanton corrections to the Kahler potential. 
For simplicity we set $\gamma_{S,T}$ to zero. 
Specific string models determine $\theta$, $\triangle$ and $\omega$. 
We leave these arbitrary for the moment.

From Eqs. (22) and (23) one has
\begin{equation}
{m_0^2\over m_{1/2}^2}={1\over3}[1+\triangle\cot^2\theta]
\end{equation}
From the discussion above, one saw that $m_0$ could be determined at a LC with 
error of about 2\% and $m_{1/2}$ with error of about 5\%.  
Thus using the parameters of the previous analyses ($\tilde{m}_2=120\ GeV$, $m_0=187\  GeV$) one finds
\begin{equation}
\triangle \cot^2 \theta=3.73\pm 0.25
\end{equation}
$A_0$ can be related to low energy parameters by the RGE yielding the relation $A_0=A_R/D_0-2.20\ m_{1/2}$ and for example chosing 
$A_t=285\ GeV$, $\tan\beta=5$ one finds $A_0/m_{1/2}= -1.539\pm 0.047$  
which yields using Eqs.(20,22):
\begin{equation}
\omega \cot \theta = 0.539 \pm 0.047
\end{equation}
Specific Calabi-Yau compactifications determine $\triangle$ and $\omega$.
Thus Eqs. (24,25) allow for two experimental determinations of $\theta$. 
We consider two models.

The value of $\triangle$ and $\omega$ can be calculated in the large Calabi-Yau 
radius limit \cite{klem}.
Thus for $Re T =5$, the one modulus models \cite{can} give average values of
\cite{kim} $\triangle \cong 0.40$, $\omega \cong 0.17$.
Then Eqs.(24) and (25) yield $\mid \cot\theta \mid = 3.05\pm0.14$; 
$ \cot\theta  = 3.17\pm0.28$.
We see for the above parameters, that these two values are consistent.
One can then determine the gravitino mass from (22) yielding 
$m_{3/2}=(276\pm 18)\ GeV$. 
The model can then be subject to other experimental tests for universality 
at the string scale, as described above. Both $\theta$ and $m_{3/2}$ are aspects 
of string supersymmetry breaking.
Thus it is possible to experimentally verify at a LC what these predictions would 
be for this model, once an understanding of supersymmetry breaking in string 
theory is obtained.

As a second model one choses the value $Im T=1/4$ which maximizes $\triangle$. 
Then $Re\ T=5$ yields \cite{kim} $\triangle=1.62$, $|\omega   |=0.64$.
Eqs. (24) and (25) would then yield for the above parameters 
$|\cot \theta|=1.516\pm 0.071$ and $|\cot \theta|=0.842\pm 0.073$ showing 
that it would be experimentally possible to rule out this string model.

\vspace{.2in}

\noindent
{\bf CONCLUSIONS}

\vspace{.2in}

LEP has allowed for precision test of the Standard Model, i.e.  physics 
$ {\tiny \begin{array}{l}  <\\ \sim \end{array}} 100 GeV$, 
but also from grand unification analyses,  it has been possible 
to probe physics up to the GUT scale.
Linear colliders and the LHC will be able to unravel the new physics that lies in 
the $TeV$ region above the electroweak scale. 
In addition, we have seen here that linear colliders will be able to probe physics 
up to the Planck scale and test assumptions made in the post-GUT domain. 
We have illustrated this here with three classes of models:

\vspace{.1in}

$\bullet$ Supergravity models with universal soft breaking at $M_{str}$. 
The predicted loss of universality at $M_G$ could be well measured, different 
gauge groups distinguished [i.e. $SU(5)$, $SO(10)$], and the value of 
$M_{str}$ measured.

$\bullet$  Models with horizontal symmetry, e.g. $SU(2)_H$.
The general $SU(2)_H$ symmetry is easily observable, and the soft breaking parameters of such models well measured.

$\bullet$ Simple Calabi-Yau string models.
Different compactifications could be distinguished and explicit string 
quantities (e.g. nature of goldstino, value of the gravitino mass $m_{3/2}$) 
can be well measured.

Thus, linear colliders are potentially very powerful experimental tools for 
unraveling physics at the Planck scale.

\end {document}